\documentclass[sigplan,9pt,screen,natbib,nonacm]{acmart}
\setcopyright{none}
\usepackage{hyphenat}

\usepackage{mdwtab}
\usepackage{mdwlist}
\usepackage{amsmath}
\usepackage{comment}
\usepackage{microtype}
\usepackage{url}
\usepackage{tikz}
\usepackage{listings}

\lstdefinestyle{sOcaml}{language=[Objective]Caml,
  literate={+}{{$+$}}1 {/}{{$/$}}1 
           {=}{{$=$}}1
           {>}{{$>$}}1 {<}{{$<$}}1
           {<>}{$\not=$}1
           {->}{{$\rightarrow$}}2 {>=}{{$\geq$}}2 {<-}{{$\leftarrow$}}2
           {<=}{{$\leq$}}2
           {==>}{{$\mapsto$}}2
           {|}{{$\mid$}}1
           {|>}{{$\blacktriangleright$}}1
           {>>}{{$\rhd$}}1
           {'a}{$\alpha$}1
           {'b}{$\beta$}1
           {'c}{$\gamma$}1
           {'e}{$\epsilon$}1
           {'i}{$\iota$}1
           {...}{\ldots}2
           {\#\#+}{\color{red}}1
           {\#\#-}{\color{black}}1
           {\#\#\#}{{$\leadsto$}}3
}
\lstnewenvironment{code}{\lstset{basicstyle={\linespread{1.02}\sffamily\normalsize}}}{}

\lstMakeShortInline[columns=fullflexible]|

\begin{document}

\title{Demo: New View on Plasma Fractals}
\subtitle{From the High Point of Array Languages}

\author{Oleg Kiselyov}
\orcid{0000-0002-2570-2186}
\affiliation{%
  \institution{Tohoku University}
  \country{Japan}}
\email{oleg@okmij.org}

\author{Toshihiro Nakayama}
\affiliation{%
  \institution{Tohoku University}
  \country{Japan}}
\email{nakayama.toshihiro.t6@dc.tohoku.ac.jp}

\begin{abstract}
Plasma fractals is a technique to generate random and realistic
clouds, textures and terrains~-- traditionally using recursive
subdivision. We demonstrate a new approach, based on iterative
expansion. It gives a family of algorithms that includes
the standard square-diamond algorithm and offers various interesting
ways of extending it, and hence generating nicer pictures.  The
approach came about from exploring plasma fractals from the point of
view of an array language (which we implemented as an embedded DSL in
OCaml)~-- that is, from the perspective of declaring whole image
transformations rather than fiddling with individual pixels.
\end{abstract}
\maketitle

\section{Summary}
\label{s:summary}
 
Plasma fractals \cite{fournier82} are widely used to simulate clouds,
generate textures and build height-maps for realistic terrains.

We present a new approach to generating plasma fractals, generalizing
the standard square-diamond algorithm. The approach leads to a family
of algorithms which we have just started to explore. It came about as
we were contemplating realizations of plasma fractals in an array
language~-- the kind of language which discourages fiddling with individual
pixels and forces us to think of transformations on images as
a whole.

The bird-eye view afforded by array languages made us realize that
plasma fractals emerge from repeated noisy image expansion, with
appropriately scaled noise~-- in marked contrast with the conventional
view as recursive subdivision. Our algorithms are not recursive but
iterative.

Along with plasma fractals we demonstrate our array programming
language~-- in the spirit of APL but implemented as an embedded DSL in
OCaml and typed. Thanks to the
rich host language, we enjoy modularity, expressivity, standard
library and powerful abstractions while avoiding puzzling and error-prone
behavior due to implicit conversions and pervasive overloading typical
of array languages.

The complete, self-contained code is available at
\url{https://okmij.org/ftp/image/ArrayL/}.

\section{Plasma Fractals as Recursive Subdivision}

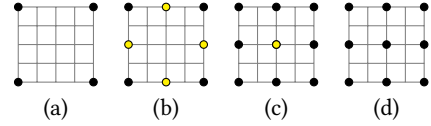
\begin{figure}
\begin{tabular}[C]{cccc}
\begin{tikzpicture}[scale=0.25]
\draw[step=1,gray,thin] (0,0) grid (4,4);
\foreach \x in {0,4}
\foreach \y in {0,4}
  \filldraw[fill=black] (\x,\y) circle [radius=0.2]; 
\end{tikzpicture}
&
\begin{tikzpicture}[scale=0.25]
\draw[step=1,gray,thin] (0,0) grid (4,4);
\foreach \x in {0,4}
\foreach \y in {0,4}
  \filldraw[fill=black] (\x,\y) circle [radius=0.2]; 
\foreach \y in {0,4}
  \filldraw[fill=yellow] (2,\y) circle [radius=0.2]; 
\foreach \x in {0,4}
  \filldraw[fill=yellow] (\x,2) circle [radius=0.2]; 
\end{tikzpicture}
&
\begin{tikzpicture}[scale=0.25]
\draw[step=1,gray,thin] (0,0) grid (4,4);
\foreach \x in {0,2,4}
\foreach \y in {0,2,4}
  \filldraw[fill=black] (\x,\y) circle [radius=0.2]; 
\filldraw[fill=yellow] (2,2) circle [radius=0.2]; 
\end{tikzpicture}
&
\begin{tikzpicture}[scale=0.25]
\draw[step=1,gray,thin] (0,0) grid (4,4);
\foreach \x in {0,2,4}
\foreach \y in {0,2,4}
  \filldraw[fill=black] (\x,\y) circle [radius=0.2]; 
\end{tikzpicture}
\\
(a) & (b) & (c) & (d)
\end{tabular}
\caption{Midpoint displacement algorithm}
\label{f:mid}
\end{figure}

Conventionally, plasma fractals are presented as recursive
subdivision, following \cite{miller86}. In the simplest midpoint
displacement algorithm, we start with the initially empty image of the
target size and seed its four corners: black circles in
Fig.~\ref{f:mid}(a). Midpoints along the sides (yellow circles in
Fig.~\ref{f:mid}(b)) are filled with the averages of the two closest
corners~-- plus a random displacement. The center point
(Fig.~\ref{f:mid}(c)) is the average of the four corners, plus a
random displacement. As the result, the four sub-squares
(Fig.~\ref{f:mid}(d)) have their corners filled, and the procedure is
recursively applied to them (with a scaled down random
displacement)~-- until the entire image is filled.

The square-diamond algorithm is an elaboration to avoid line artifacts of the
simple algorithm.

\section{Plasma Fractal = Expansion + Noise}
\label{s:expansion}

As we were pondering how to implement plasma fractals in an array
language (see \S\ref{s:array-lang}) and avoid concerning ourselves with
filling-in of individual pixels, we have come across a new presentation.
Whereas the convention talks about recursive subdivision, we think of
progressive (iterative) enlargement.

\begin{figure}
\begin{tabular}[C]{c@{\hskip 3em}c}
\begin{minipage}{0.15\columnwidth}
\begin{tikzpicture}[scale=0.4]
\draw[step=1,gray,thin] (0,0) grid (1,1);
\foreach \x in {0,1}
\foreach \y in {0,1}
  \filldraw[fill=black] (\x,\y) circle [radius=0.2]; 
\end{tikzpicture}
\vspace{2em}
\end{minipage}
&
\begin{tikzpicture}[scale=0.4]
\draw[step=1,gray,thin] (0,0) grid (2,2);
\foreach \x in {0,1,2}
\foreach \y in {0,1,2}
  \filldraw[fill=yellow] (\x,\y) circle [radius=0.2]; 
\foreach \x in {0,2}
\foreach \y in {0,2}
  \filldraw[fill=black] (\x,\y) circle [radius=0.2]; 
\end{tikzpicture}
\\
(a) & (b)
\end{tabular}
\caption{Expansion Algorithm}
\label{f:expansion}
\end{figure}
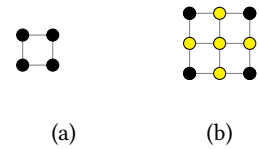

We start with a $2\times2$ seed picture, Fig.~\ref{f:expansion}(a), and
expand it: Fig.~\ref{f:expansion}(b). The yellow circles
represent `new pixels', whose values are determined by resampling
the original image, with the added random noise. With bilinear
interpolation as resampling, the side pixels are the averages of their
two side neighbors, and the middle is the
average of its four neighbors. The result is hence equivalent to
(a scaled down) Fig.~\ref{f:mid}(d). Repeating the expansion gives
the image of the target size.

Thinking in terms of images rather than pixels (as array languages
encourage us to do \cite{McIntyre91,life-with-AL}), the noisy
expansion step is the standard image enlargement to which a noise
image is added. The enlargement (upsampling) can be represented as a
2D-convolution. For bilinear interpolation upsampling, the convolution
kernel is the $2\times 2$ array: is the outer product of |[1; 1]/2|
with itself.  Interestingly, the square-diamond algorithm can also be
represented as the repeated scaling/convolution+noise, using a
particular convolution kernel.  Fig.~\ref{f:bicubic-fractals} shows
sample plasma fractals generated with bicubic interpolation
upsampling~-- the new algorithm that improves on square-diamond and
produces images with fewer artifacts.

\section{Array Language}
\label{s:array-lang}

What made the exploration of plasma fractals easier is an array
language~-- in our case, one embedded in OCaml. It gave a playground to
quickly try various algorithms and see their results.

Our language is greatly inspired by APL. Unlike APL, J, K, etc.,
however, it has the conventional programming language syntax (actually
that of OCaml): expressions may span as many lines as needed to nicely
display them, highlight their structure and attach comments; one may
locally name (sub)expressions for readability; parsing/understanding
requires less look-ahead.\footnote{APL parsing requires unlimited
  look-ahead (or, more precisely, look-behind, since APL is parsed
  right-to-left): \url{http://dfns.dyalog.com/n_parse.htm}.}
There are further, less noticeable but just
as important differences from APL: the flow of data is left-to-right
(as common in electrical engineering and signal processing) rather
than right-to-left. Not everything is an array: there are also
numbers, pairs, strings, booleans, and all other OCaml data
types. There are types. Implicit conversions, padding, overloading,
slicing are eliminated to minimize surprises and subtle errors,
especially due to typos. For example, a conversion from |int| to
|float| has to be notated explicitly, using OCaml's standard
|float_of_int|. That may become cumbersome~-- but OCaml's local module
\textsf{open}, local definitions and other abstractions help. The
whole OCaml and its libraries are available for syntax sweetening, in
the manner advocated in \cite{EffNoMonads}.

Still our language is an array language, distinguishing
array shape and contents, relying on compositions and array
arithmetic; tacit programming (a.k.a. point-free style) is also
available, with the set of common combinators.
Our arrays are defined as 
\begin{code}
type ('i,'a) arr = Arr of 'i * ('i -> 'a)
\end{code}
literally as the combination of shape, or index domain (represented by 
the type
|'i|) and content: the function from an index to an element. 
For images, the index domain type is the pair of integers
\begin{code}
type d2 = int * int
\end{code}
and the shape is the pair of the largest row and column indices.
In this representation, operations on arrays are automatically
fused and copying is
avoided. In fact, no arrays are allocated unless explicitly
requested, by calling |materialize2|.

As an example, the noisy enlargement step of the plasma fractal
algorithm from \S\ref{s:expansion} is implemented as
\begin{code}
let noise (Arr (d,_)) = Arr (d, fun (i,j) -> 
       if i land 1 = 0 && j land 1 = 0 then 0 else rand ())

let expander (scaler: ((d2,int) arr -> (d2,int) arr)) (nsf: float)
 : (d2,int) arr -> (d2,int) arr = 
  map (fimul nsf) >> scaler >> fun m2 ->
  zip_with (+) m2 (noise m2) |> materialize2 0
\end{code}
where \lstinline{|>} is left-to-right application and
\lstinline{>>} is left-to-right composition. The type annotations are
optional and given for clarity.
The argument |scaler| is an enlargement function, such as bilinear,
bicubic or square-diamond convolutional upsampling. The argument
|nsf| is noise scaling related to the fractal
dimension. For plasma fractals, it should be around |1.1-2.2|.

\begin{figure}[h]
\includegraphics[scale=0.5]{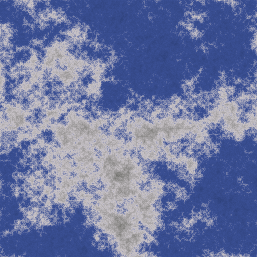}

\medskip
\includegraphics[scale=0.5]{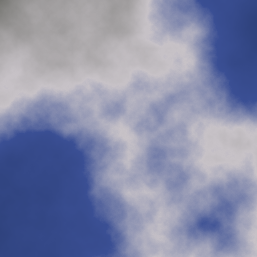}

\caption{Generated plasma fractals (resampling by bicubic interpolation)}
\label{f:bicubic-fractals}
\end{figure}

Fig.~\ref{f:bicubic-fractals} (top) is the result of
\begin{code}
let m0 = of_array [|4;4;4;4|] |> rho2 (2,2) in
m0 |> ntimes 8 (expander scale_twice_bc 1.2)
\end{code}
The bottom image on the figure is obtained with the |nsf| parameter set to
|2.0|.

\section{From Now On}

The new family of plasma fractal algorithms is vast: any image
expansion algorithm\footnote{see \url{https://en.wikipedia.org/wiki/Resampling_(bitmap)}
for a good list} instantly gives a plasma fractal algorithm. 
We have just begun to explore this landscape.

\bibliographystyle{plainnat}
\bibliography{refs.bib}

\begin{thebibliography}{5}
\providecommand{\natexlab}[1]{#1}
\providecommand{\url}[1]{\texttt{#1}}
\expandafter\ifx\csname urlstyle\endcsname\relax
  \providecommand{\doi}[1]{doi: #1}\else
  \providecommand{\doi}{doi: \begingroup \urlstyle{rm}\Url}\fi

\bibitem[Fournier et~al.(1982)Fournier, Fussell, and Carpenter]{fournier82}
Alain Fournier, Donald~S. Fussell, and Loren~C. Carpenter.
\newblock Computer rendering of stochastic models.
\newblock \emph{Communications of the ACM}, 25\penalty0 (6):\penalty0 371--384,
  June 1982.
\newblock \doi{10.1145/358523.358553}.

\bibitem[Kiselyov(2019)]{EffNoMonads}
Oleg Kiselyov.
\newblock Effects without monads: {N}on-determinism~-- back to the {M}eta
  {L}anguage.
\newblock \emph{EPTCS}, 294:\penalty0 15--40, 2019.
\newblock \doi{10.4204/EPTCS.294.2}.

\bibitem[McIntyre(1991)]{McIntyre91}
Donald~B. McIntyre.
\newblock Language as an intellectual tool: From hieroglyphics to {APL}.
\newblock \emph{IBM Systems Journal}, 30\penalty0 (4):\penalty0 554--581, 1991.
\newblock \doi{10.1147/sj.304.0554}.

\bibitem[Miller(1986)]{miller86}
Gavin S.~P. Miller.
\newblock The definition and rendering of terrain maps.
\newblock In David~C. Evans and Rusell~J. Athay, editors, \emph{Proceedings of
  the 13th Annual Conference on Computer Graphics and Interactive Techniques,
  SIGGRAPH 1986, Dallas, Texas, USA, August 18-22, 1986}, volume~20, pages
  39--48, August 1986.
\newblock \doi{10.1145/15922.15890}.

\bibitem[Smillie(2005)]{life-with-AL}
Keith Smillie.
\newblock My life with array languages.
\newblock October 2005.

\end{thebibliography}
\end{document}